\documentclass[lettersize,journal]{IEEEtran}
\usepackage{amsmath,amsfonts}
\usepackage{algorithmic}
\usepackage{algorithm}
\usepackage{array}
\usepackage[caption=false,font=normalsize,labelfont=sf,textfont=sf]{subfig}
\usepackage{textcomp}
\usepackage{stfloats}
\usepackage{url}
\usepackage{verbatim}
\usepackage{graphicx}
\usepackage{cite}
\begin{document}

\title{A Novel Gradient Descent Least Squares (GDLS) Algorithm for Efficient SMV Gridless Line Spectrum Estimation with Applications in Tomographic SAR Imaging}

\author{
	Ruizhe Shi,
	Zhe Zhang,~\IEEEmembership{Member,~IEEE,}
	Xiaolan Qiu,~\IEEEmembership{Senior Member,~IEEE,}
	and Chibiao Ding
%
%
%
%
%
%
\thanks{Part of this work was supported by the NSFC grant \#61991421, \#61991420 and \#62022082, AIRCAS grant ``Structured signal high efficiency sensing theory and its application in microwave imaging'', and ``Suzhouo Gusu Leading Talents Project'' grant \#ZXL2022381.}
\thanks{Part of this work was presented on the 2021 IEEE International Geoscience and Remote Sensing Symposium (IGARSS 2021).}

\thanks{R. Shi, Z. Zhang, X. Qiu and C. Ding are with Aerospace Information Research Institute, Chinese Academy of Sciences (AIRCAS) and School of Electronic, Electrical and Communication Engineering, University of Chinese Academy of Sciences, Beijing 100190, China. }%
\thanks{R. Shi, Z. Zhang and X. Qiu are also with Suzhou Aerospace Information Research Institute and Suzhou Key Laboratory of Intelligent Aerospace Big Data Application Technology, Suzhou, Jiangsu 215123, China. }%
\thanks{R. Shi, X. Qiu and C. Ding are also with CAS Key Laboratory of Technology in Geo-spatial Information Processing and Application System, Beijing 100190, China. }%

\thanks{Corresponding author: Zhe Zhang (Email: zhangzhe01@aircas.ac.cn)}
}

\markboth{Journal of \LaTeX\ Class Files,~Vol.~14, No.~8, August~2021}%
{Shell \MakeLowercase{\textit{et al.}}: A Sample Article Using IEEEtran.cls for IEEE Journals}


\maketitle

\begin{abstract}
This paper presents a novel efficient method for gridless line spectrum estimation problem with single snapshot, namely the gradient descent least squares (GDLS) method. Conventional single snapshot (a.k.a. single measure vector or SMV) line spectrum estimation methods either rely on smoothing techniques that sacrifice the array aperture, or adopt the sparsity constraint and utilize compressed sensing (CS) method by defining prior grids and resulting in the off-grid problem. Recently emerged atomic norm minimization (ANM) methods achieved gridless SMV line spectrum estimation, but its computational complexity is extremely high; thus it is practically infeasible in real applications with large problem scales. Our proposed GDLS method reformulates the line spectrum estimations problem into a least squares (LS) estimation problem and solves the corresponding objective function via gradient descent algorithm in an iterative fashion with efficiency. The convergence guarantee, computational complexity, as well as performance analysis are discussed in this paper. Numerical simulations and real data experiments show that the proposed GDLS algorithm outperforms the state-of-the-art methods e.g., CS and ANM, in terms of estimation performances. It can completely avoid the off-grid problem, and its computational complexity is significantly lower than ANM. Our method has been tested in tomographic SAR (TomoSAR) imaging applications via simulated and real experiment data. Results show great potential of the proposed method in terms of better cloud point performance and eliminating the gridding effect. 
\end{abstract}


\section{Introduction}
\IEEEPARstart{L}ine  spectrum estimation is an important problem in the field of signal processing with broad applications, such as direction of arrival (DOA) \cite{ref1}, radar signal processing and imaging \cite{ref2}, wireless communication \cite{ref3}, seismology \cite{ref4}, medical imaging \cite{ref5}, etc.  Usually, line spectrum estimation problems deal with a linear mixture of sinusoidal (exponential) signals and our main goal is to recover the frequencies and amplitudes from it with efficiency under some constraints. 

Most classical line spectrum estimation algorithms are based on feature subspace decomposition, usually known as subspace methods. Starting with the sample statistics, some use the noise subspaces such as the eigenvector method \cite{ref6} and the multiple signal classification (MUSIC) algorithm \cite{ref7}; others use the signal subspace, such as the ESPRIT algorithm \cite{ref8} and the TLS-ESPRIT algorithm \cite{ref9}. These techniques could achieve super-resolution (which means no pre-defined searching grid), and work effectively with multiple snapshots (a.k.a. multiple measurement vector or MMV) to well approximate the signal sample statistics. Additionally, they usually require the sources to be uncorrelated. In order to improve the availability of subspace methods in single snapshot (or single measurement vector, SMV) and/or correlated cases, spatial smoothing based techniques are introduced, such as the iterative spatial smoothing (ISS) algorithm \cite{ref10}, the weighted spatial smoothing (WSS) algorithm \cite{ref11}, the spatial difference smoothing (SDS) algorithm \cite{ref12} and the damped MUSIC (DMUSIC) algorithm \cite{ref37}.  With smoothing, subspace methods work for correlated and/or SMV sources at the cost of reducing the effective array aperture size and hence sacrificing the resolution. Besides, subspace methods rely on some prior known information, such as the number of sources.

Compressed sensing (CS) \cite{ref13,ref14} is a popular structure-based signal processing framework that suggests one can recover a signal from highly compressed samples if the original signal is sparse under some basis. As an effective way to reduce the number of measurements with provable performance guarantees by promoting the sparsity prior \cite{ref15}, CS has been introduced into line spectrum estimation problems to tackle the issues of subspace methods \cite{ref16,ref17,ref18,ref19}. Compared with subspace methods, CS can easily deal with correlated sources with even single snapshot, without the expense of array aperture. However, CS works under a gridding assumption by dividing the frequency domain into pre-defined virtual grids. Hence, conventional CS approaches suffer from limited resolution, and the mismatch between true signal frequencies and the pre-defined grids could cause the off-grid effect, which will result in the power leakage into neighboring grids, reducing the sparsity and affects the estimation performance. This is a massive disadvantage in performance compared with traditional subspace methods. 

To mitigate the off-grid issue in CS, gridless CS approaches has been developed. Traditional gridless CS approaches usually treat the off-grid effect as a perturbation from the ideal grid and aim to refine it via iterations \cite{ref20,ref21}. Alternatively, the newly emerged atomic norm minimization (ANM) method is proposed as a gridless version of the conventional CS to recover the off-grid sinusoidal components from compressive measurements \cite{ref22}. Compared with CS, which relies on the sparsity structure, ANM exploits the Vandermonde structure of the signal to attain off-the-grid estimation at super-resolution and performs well with a single snapshot, correlate signal, and off-grid statement. The application of ANM in line spectrum estimation is successful and it has been extended into 2-D cases \cite{ref23} ,  with applications in various fields such as DOA estimation, channel estimation and radar imaging \cite{ref24,ref25,ref26}. The main issue of ANM is its computational complexity because ANM problems are usually solved by semidefinite programming (SDP) method, which is believed be highly expensive in computation and makes it infeasible when the problem scale goes large. Some approaches have been proposed to accelerate the ANM computation \cite{ref27}, but the computational complexity is still limited by the SDP framework.

The maximum likelihood estimation (MLE) is a kind of classical estimation algorithms, which provide the most accurate frequency estimates under a given distribution \cite{ref38}. Nonetheless, because the solution is a multivariate nonlinear maximization problem, its high computational complexity affects the use of the algorithm. One solution is to decompose the objective function into several one-dimensional problems, such as alternating projection \cite{ref34} or the IQML (iterative quadratic ML) \cite{ref39}. There are also some algorithms using artificial intelligence like simulated annealing to solve the function\cite{ref35}. However, all the above algorithms still require a lot of computation, which is unacceptable in practical application. For real applications such as tomographic SAR imaging, oftentimes only single snapshot scenario needs to be considered. Therefore, the derivative of the objective function is analytical, which makes the gradient optimization method possible.

In this paper, we propose a novel single snapshot gridless line spectrum estimation method by approximating the complex amplitude of sinusoids via least square (LS) estimation and solving the modified minimum norm problem with gradient descent, namely the gradient descent least squares (GDLS) algorithm. By formulating the line spectrum estimation into LS estimation form and carefully choosing the initial value to guarantee the convergence, the proposed GDLS algorithm does not suffer from the off-grid problem and works effectively with single snapshot without sacrificing performances in terms of resolution, noise robustness and accuracy. Compared to ANM approaches, its computational complexity is markedly reduced from $\mathcal{O}(N^{3.5} )$ to $\mathcal{O}(N^2 )$, which is a significant jump in efficiency.

Tomographic synthetic aperture radar (TomoSAR) imaging is an important three-dimensional microwave remote sensing technique which forms a synthetic aperture along the elevation direction to achieve resolving ability of multiple targets within one range-azimuth resolution cell \cite{ref30,ref31,ref32,ref33}. TomoSAR enables traditional radar imaging to have three-dimensional imaging ability and has attracted great interested in the recent years. TomoSAR has a very typical line spectrum estimation model along the elevation model, and traditionally people use canonical subspace methods or compressed sensing based methods to deal with it \cite{refadd1,refadd2}, so it is natural to exploit our proposed GDLS method in TomoSAR applications, aiming at benefits including gridless focusing and faster convergence.

The rest of this paper is organized as follows. In Section 2, the problem formulation is introduced. The GDLS algorithm is proposed in Section 3. Further issues, including comparison with ML estimation, convergence analysis, noise robustness and complexity analysis are discussed in Section 4. The numerical results in Section 5 validate the effectiveness and efficiency of the proposed GDLS method. We applied the proposed method in TomoSAR imaging in Section 6 and validated its performance superiority via simulated and real data, followed by conclusions in Section 7.

\section{Problem setting}

Consider a line spectrum estimation problem with $L$ sources given in the form 
\begin{equation}\label{key1}
	x(t)=\sum_{l=1}^{L}c_le^{j2\pi f_lt},
\end{equation} 
where $f_l \in [0, 1)$ is the normalized frequency of the $l$-th source, $c_i\in \mathbb{C}$ is the complex amplitude of the $l$-th source and $t$ is the time. Suppose we sample at the time domain in a uniform manner as

\begin{equation}\label{key2}
	t=nT,n=1,...,N,
\end{equation}
where $N$ is the number of samples and $T$ is the sample interval, \eqref{key1} can be expressed as
\begin{equation}\label{key3}
	x(n)=\sum_{l=1}^{L}c_le^{j2\pi nf_l}.
\end{equation}

Equation \eqref{key3} can be written in vector form as a complex weighted mixture of sinusoids with distinct frequencies $f_l \in [0, 1)$
\begin{equation}\label{key4}
	\mathbf{x}=\sum_{l=1}^{L}c_l\mathbf{a}(f_l),
\end{equation}
where $\mathbf{a}(f_l )$ is a vector determined by $f_l$,
\begin{equation}\label{key5}
	\mathbf{a}(f_l)=[1, e^{j2\pi f_l}, \cdots, e^{j2\pi (N-1)f_l}]^{\mathrm{T}},
\end{equation}
which is usually referred as array steering vector in array signal processing.

Our goal is to recover those sinusoidal components, especially the frequencies $\{\mathbf{f}\in \mathbb{R}^{L}|f_l\in[0,1)\}$, from the measurements $\mathbf{x}$, or its noisy version $\mathbf{y}$ with additive Gaussian white noise $\mathbf{w}$
\begin{equation}\label{key6}
	\mathbf{y}=\mathbf{x}+\mathbf{w}.
\end{equation}

Such a problem arises in many applications while the physical meanings of digital frequencies $f_l$ may vary. We give two examples.
\subsection{DOA estimation}

Shown as Fig 1, a unitary linear array (ULA) can be used to estimate the direction of arrival \cite{ref28,ref29}. Under far-field conditions, the signal wavefronts are arrived in parallel. Assume the array elements spacing is $d$ and the signal received by the first array element is $x_1$, then the signal received by the $n$-th element can be written as
\begin{equation}\label{key7}
	x_n = x_1\cdot \exp\left\{j\frac{2\pi (n-1)d}{\lambda}\cos\theta\right\},
\end{equation}
where $\theta$ is the arrival angle. In a compact form as
\begin{equation}\label{key8}
	\mathbf{x}=[x_0, \cdots , x_{N-1}]^{\mathrm{T}},
\end{equation}
which is an array steering vector modulated by amplitude $x_0$, and the digital frequency is
\begin{equation}\label{key9}
	f_l=\frac{d}{\lambda}\cos\theta.
\end{equation}

If we have more than one sources, \eqref{key8} turns out to be the form as \eqref{key4}.

\begin{figure}[h]
	
	\centering
	\includegraphics[width=2.5in]{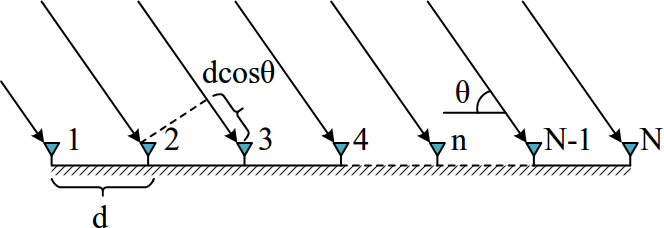}
	\caption{DOA estimation geometry with ULA}
	\label{fig1}
	
\end{figure}

\subsection{TomoSAR imaging}
Fig 2 shows the model of TomoSAR imaging. Consider the same range-azimuth cell was observed by a radar for multiple times. As shown in reference \cite{ref30}, on each radar image which contains the same target scene from different observations, the echo phase delay of the same ground object is different. After the first-order Taylor approximation, the phase differences of the target between different observations are proportional to the height difference $s$ along the elevation axis, which is orthogonal to the direction of arrival\cite{ref30}. 
\begin{equation}\label{keyadd1}
	x_n=\int_{\triangle s}\gamma(s)\exp(j2\pi \xi_n s)\mathrm{d}s ,
\end{equation}
\begin{equation}\label{keyadd2}
	\xi_n=2b_{n} /\lambda R_0,
\end{equation}
where $\gamma(s)$ represents the reflectivity function along elevation $s$, $\lambda$ is the wavelength of the radar, and $R_0$ is the range of the scatterer from the sensor.

Assuming that the antenna array is evenly distributed, the baseline length of the $n$-th antenna $b_{n}$ is $n-1$ multiplies with the baseline spacing $b$,
\begin{equation}\label{keyadd3}
	b_{n}=(n-1)b.
\end{equation}
Thus, the target echo received by the $n$-th observation can be written as
\begin{equation}\label{key10}
	x_n=x_1\cdot e^{j2\pi (n-1)f},
\end{equation}
\begin{equation}\label{key11}
	f=\dfrac{2bs}{\lambda R_0},
\end{equation}
where $f\in[0,1)$ is the normalized digital frequency  determined by the elevation $s$. It also follows a typical line spectrum estimation form.
\begin{figure}[!h]
	\centering
	\includegraphics[width=2.5in]{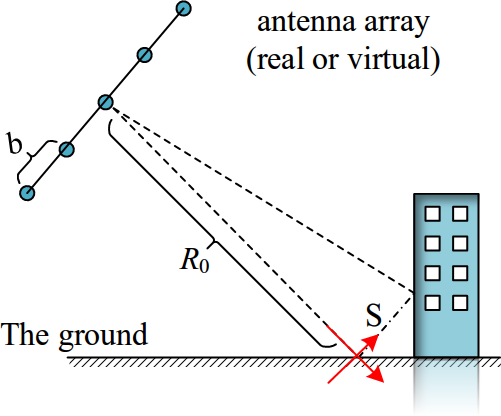}
	\caption{TomoSAR imaging geometry. The coordinate s is referred to as elevation.}
	\label{fig2}
\end{figure}

\section{Proposed method }

\subsection{Objective function of least squares}
Recall the problem model of \eqref{key6} as
\begin{equation}\label{key12}
	\mathbf{y}=\mathbf{A}\mathbf{c}+\mathbf{w},
\end{equation}
where
\begin{equation}\label{key13}
	\mathbf{A}=[\mathbf{a}(f_1),\cdots,\mathbf{a}(f_L)],
\end{equation}
is the matrix composed by the array steering vectors (also known as the array manifold) and
\begin{equation}\label{key14}
	\mathbf{c}={[c_1,\cdots,c_L]}^{\mathrm{T}},
\end{equation}
is the column vector composed of $c_l$. 

In order to estimate the digital frequencies $f_l$, we firstly deal with the amplitude $\mathbf{c}$ which is also unknown. When the frequencies $f_l$ are known, the estimation of the amplitude coefficient $c_l$  can be described as a solution to overdetermined equations in a least squares (LS) manner as 
\begin{equation}\label{key15}
	\hat{\mathbf{c}}=\arg  \mathop{\min}_{\mathbf{c}} {\Vert \mathbf{y}-\mathbf{A}\mathbf{c} \Vert}^{2}  .
\end{equation}

The solution to this LS problem is unique, 
\begin{equation}\label{key16}
	\hat{\mathbf{c}}=\mathbf{A}^{\mathrm{+}}\mathbf{y}={(\mathbf{A}^{\mathrm{H}}\mathbf{A})^{-1}}\mathbf{A}^{\mathrm{H}}\mathbf{y},
\end{equation}
so we can substitute it back into the observation error of the signal, obtaining a cost function only related to the frequencies $f_l$. This is the expression of the objective function of proposed gradient descent least square (GDLS) problem:
\begin{equation}\label{key17}
	f_l=\arg \mathop{\min}_{f_l}\Vert\mathbf{y}-\mathbf{A}\mathbf{A}^{\mathrm{+}}\mathbf{y}\Vert^2.
\end{equation}

\subsection{Solving gradient descent}
The cost equation in \eqref{key17} describes an $L$-order search problem, which is usually solved by first-order algorithms, such as the gradient descent method. The details of gradient descent are derived as follows.
The gradient of the objective function can be calculated by the following formula:
\begin{equation}\label{key18}
	T(\mathbf{f})=\Vert\mathbf{y}-\mathbf{A}\mathbf{A}^{\mathrm{+}}\mathbf{y}\Vert^2=\mathbf{y}^{\mathrm{H}}\mathbf{y}-\mathbf{y}^{\mathrm{H}}\mathbf{A}\mathbf{A}^{\mathrm{+}}\mathbf{y},
\end{equation}
\begin{equation}\label{key19}
	\dfrac{\partial T}{\partial f_l}=\mathbf{y}^{\mathrm{H}}\left(\mathbf{Q}+\mathbf{Q}^{\mathrm{H}}\right)\mathbf{y},
\end{equation}
where
\begin{equation}\label{key20}
	\mathbf{Q}=\left(\mathbf{I}-\mathbf{A}\mathbf{A}^{\mathrm{+}}\right)\dfrac{\partial \mathbf{A}}{\partial f_l}\mathbf{A}^{\mathrm{+}},
\end{equation}
\begin{equation}\label{key21}
	\dfrac{\partial \mathbf{A}}{\partial f_l}=[\mathbf{0},\cdots,\dfrac{\partial \mathbf{a}(f_l)}{\partial f_l},\cdots,\mathbf{0}],
\end{equation}
\begin{equation}\label{key22}
	\dfrac{\partial \mathbf{a}(f_l)}{\partial f_l}=[0,j2\pi e^{j2\pi f_l},\cdots,j2(N-1)\pi e^{j(N-1)2\pi f_l}]^{\mathrm{T}}
\end{equation}
\begin{equation}\label{key23}
	\mathbf{f}=[f_1,\cdots,f_L]^{\mathrm{T}}.
\end{equation}

\subsection{Iteration initial value}
It should be noted that the objective function \eqref{key17} is nonconvex. It has multiple local minima points in the search domain, and the gradient changes rapidly makes the gradient descent difficult to converge to the global optima. Tacking this difficulty, we provide an effective method by carefully selecting the initial value.

The estimation of compressed sensing methods suffers from the gridding effect, but in most cases, they are still close to the real values. Besides, CS methods, especially the greedy algorithms like OMP algorithm \cite{ref16} are relatively efficient. Hence, we suggest that use the rough estimations of the OMP algorithm as the initial value of iteration, and use the GDLS as a refining step after OMP by doing small step length iterations.

In most cases, our proposed method converges properly to the correct global optima with efficiency. The detailed steps of the whole algorithm are shown in Algorithm 1. The algorithm inputs the observation signal and a priori estimated sparsity, and outputs the estimation of line spectrum frequencies.

\begin{algorithm}[H]
	\caption{The proposed GDLS algorithm.}\label{alg:alg1}
	\begin{algorithmic}
		\STATE 
		\STATE {\textbf{Input}} {observation $\mathbf{x}$; sparsity $L$;}
		\STATE  \textbf{Initialize} $j =0$, $\mathbf{f}^0=$OMP$(\mathbf{x})$ ,$\alpha$
		\STATE \textbf{While} not meet the stop condition \textbf{do}
		\STATE \hspace{0.5cm}$\mbox{Grad}\leftarrow-\frac{\partial}{\partial \mathbf{f}}\left(\mathbf{x}^{\mathrm{H}}\mathbf{A}\mathbf{A}^{\mathrm{+}}\mathbf{x}\right)|_{\mathbf{f} =\mathbf{f}^j }$
		\STATE \hspace{0.5cm}$\mathbf{f}^{j+1}\leftarrow\mathbf{f}^j+\alpha\frac{\mbox{Grad}}{\Vert\mbox{Grad}\Vert}$
		\STATE \hspace{0.5cm} $j\leftarrow j+1$
		\STATE \textbf{end} 
		\STATE {\textbf{Output}} estimated frequencies $ \mathbf{f}^j  $
	
	\end{algorithmic}
	\label{alg1}
\end{algorithm}

\section{Performance analysis and discussions}
Some issues of the proposed method as well as performance metrics are discussed in this section, including convergence analysis, computational complexity, number of sources, etc. 

\subsection{Compared with the ML estimation}
Towards the line spectrum estimation problem, classical maximal likelihood (ML) based algorithms share some similarities with our proposed method and the two algorithms are discussed in this section.

The existing ML algorithms \cite{ref34,ref35} mainly focus on the MMV case with multiple snapshots, and the basic idea is to estimate the parameters according to the sample set obtained from multiple observations and calculate the covariance matrix by sample averaging. The optimization problem of the ML method is
\begin{equation}\label{key24}
	\mathop{\max}_{\mathbf{f}}=tr(\mathbf{P}_{\mathbf{A}}\mathbf{R}),
\end{equation}
where $\mathbf{P}_{\mathbf{A}}$ is the projection matrix onto the space spanned by the columns of $\mathbf{A}$, and
\begin{equation}\label{key25}
	\mathbf{R}=\frac{1}{T}\sum_{t=1}^{T}\mathbf{y}_t {\mathbf{y}_t}^{\mathrm{H}}
\end{equation}
is the sample covariance matrix of the measured vectors calculated by averaging $T$ snapshots.

The relationship of ML method and our proposed GDLS method can be described as follows. We treat the observed signal $\mathbf{y}$ as one sample of a Gaussian random process parametered by mean $\mathbf{x}$, and covariance $\sigma\mathbf{I}$ where $\sigma$ is the noise level of $\mathbf{w}$ .i.e., the distribution of $\mathbf{y}$ follows
\begin{equation}\label{key26}
	\mathbf{y}\sim N(\mathbf{A}\mathbf{c},\sigma^2\mathbf{I}),
\end{equation}
the PDF of $\mathbf{y}$ is
\begin{equation}\label{key27}
	g(\mathbf{y})=\frac{1}{\left(2\pi\right)^{\frac{N}{2}}\sigma^N}\exp\left(-\frac{1}{2\sigma^2}\Vert\mathbf{y}-\mathbf{A}\mathbf{c}\Vert^2\right).
\end{equation}
Its Lagrangian is
\begin{equation}\label{key28}
	\mathrm{L}(\mathbf{f},\mathbf{c},\sigma^2;\mathbf{y})=-\frac{N}{2}\log\sigma^2-\frac{1}{2\sigma^2}\Vert\mathbf{y}-\mathbf{A}\mathbf{c}\Vert^2,
\end{equation}
and if we estimate it via ML method, we have
\begin{equation}\label{key29}
	(\hat{\mathbf{f}},\hat{\mathbf{c}})=\arg \min \Vert\mathbf{y}-\mathbf{A}\mathbf{c}\Vert^2,
\end{equation}
which is consistent with the proposed GDLS method. 

This shows that the result of our method satisfies the maximum likelihood condition and is mathematically optimal under the aforementioned conditions.

Different from the aforementioned ML algorithms, our method focuses on the SMV case with single snapshot. The objective function is mathematically differentiable, so its gradient can be explicitly expressed and the gradient descent method can be used to avoid the heavy computational load. 

In summary, our method not only has the mathematical optimal characteristics which is consistent with the maximum likelihood method, but also reduces the computational time towards the single snapshot case. This is a better trade-off between time and accuracy compared with existing methods.

\subsection{Convergence analysis and selection of initial value}
The convergence performance of the proposed method is highly relied on the selection of initial value because the objective function is non-convex. Expand the objective function and remove the constant term, we can get its equivalent form as 
\begin{equation}\label{key30}
	\begin{aligned}
		\hat{\mathbf{f}}&=\mathop{\arg\max}_{\mathbf{f}}\left(\mathbf{y}^{\mathrm{H}}\mathbf{A}\mathbf{A}^{\mathrm{+}}\mathbf{y}\right) \\
		&=\mathop{\arg\max}_{\mathbf{f}}\left(\mathbf{y}^{\mathrm{H}}\mathbf{P}_{\mathbf{A}}\mathbf{y}\right) \\
		&=\mathop{\arg\max}_{\mathbf{f}} \left<\mathbf{y},\mathbf{P}_{\mathbf{A}}\mathbf{y}\right> ,
	\end{aligned}
\end{equation}
where $\mathbf{P}_{\mathbf{A}}$ represents the projection matrix of the subspace supported by columns of matrix $\mathbf{A}$, and $< \ , \ > $ represents the inner product of two vectors. The value of the objective function can be considered as the similarity between the observed signal and its projection on the subspace spanned by the estimated parameters. 

Assuming that frequencies $f_l$ are adequately separated, the autocorrelation matrix of the observation matrix $\mathbf{A}$ is approximately the unit matrix (which means $\mathbf{A}$ is almost orthogonal), and the steering vectors $\mathbf{a}(f_l )$ corresponding to different frequencies are approximately uncorrelated. Such assumptions usually can be approximately satisfied. Besides, if the initial value $\hat{f}_l^0$ is close enough to the true value $f_l$ at the beginning, it is safe to assume that the estimated array manifold $\mathbf{A}(\hat{f}_l^K)$ is also orthogonal and $\mathbf{a}(f_l )$ is mainly projected on $\mathbf{a}(\hat{f}_l^K )$. Thus, we can further decompose the objective function and approximate it with a weighted sum of the estimated similarity of the sinusoidal signals. 

\begin{equation}\label{key31}
	\begin{aligned}
		\hat{\mathbf{f}}&=\mathop{\arg\max}_{\mathbf{f}} \left<\sum_{l=1}^{L}c_l\mathbf{a}(f_l)+\mathbf{w},\mathbf{P}_{\mathbf{A}}\left(\sum_{l=1}^{L}c_l\mathbf{a}(f_l)+\mathbf{w}\right)\right> \\
		&\approx \mathop{\arg\max}_{\mathbf{f}} \sum_{l=1}^{L}\left<c_l\mathbf{a}(f_l),\left(\mathbf{a}(\hat{f}_l)\mathbf{a}(\hat{f}_l)^{\mathrm{H}}\right)\cdot c_l\mathbf{a}(f_l)\right>\\
		&\approx \mathop{\arg\max}_{\mathbf{f}}\sum_{l=1}^{L}c_l^2\cdot \vert\mathbf{a}(\hat{f}_l)^{\mathrm{H}}\mathbf{a}(f_l)\vert^2.
	\end{aligned}
\end{equation}

For a single sinusoidal signal, the correlation between the estimated signal and the real signal is only affected by the deviation of the estimated frequency as
\begin{equation}\label{key32}
	\begin{aligned}
		\vert\mathbf{a}(\hat{f}_l)^{\mathrm{H}}\mathbf{a}(f_l)\vert&=\left\vert\sum_{n=0}^{N-1}e^{j2\pi n(f_l-\hat{f}_l)}\right\vert\\
		&=\left\vert\frac{\sin[\pi N(f_l-\hat{f}_l)]}{\sin[\pi(f_l-\hat{f}_l)]}\right\vert\\
		&=\mbox{Sad}(\pi\vert f_l-\hat{f}_l\vert,N),
	\end{aligned}
\end{equation}
where Sad$(x,n)$ is defined as
\begin{equation}\label{key33}
	\mathrm{Sad}(x,n)=\frac{\sin(nx)}{\sin(x)},
\end{equation}
and its 2-order derivate is

\begin{equation}\label{key34}
	\begin{split}
		&\frac{\partial^2\mathrm{Sad}(x,n)}{\partial x^2}\\
		&=\frac{2\sin{(n-1)x}\cos{x}-(n-1)\sin{x}\cos{(n-1)x}}{\sin^3x}.
	\end{split}
\end{equation}

Notice that,
\begin{equation}\label{key35}
	\tan{nx}\geq n\tan{x},\rm{~~~}\vert nx \vert <\pi,
\end{equation}
we can determine that the convex domain of the function Sad$(\pi\vert f_l-\hat{f}_l \vert,N)$ near $f_l$ is
\begin{equation}\label{key36}
	\vert f_l-\hat{f}_l\vert < \frac{1}{N-1}
\end{equation}

Therefore, when the deviation of the initial value meets the above condition \eqref{key36}, which is equivalent to being less than the Rayleigh resolution, the objective function is approximately convex, and the gradient descent method is feasible with large probability. In most cases, the condition can be reached if we use the output of OMP as the initial value without losing efficiency.  
\subsection{Computational complexity}
The proposed method can be mainly organized into two steps. Firstly, the initial value of iteration needs to be determined under condition \eqref{key36}, and the OMP algorithm is used in this step. The computational complexity of OMP is $O(NML)$, where $M$ means size of the dictionary set. Secondly, in the gradient descent iteration step, the complexity of each iteration is $O(N^2 L)$ because $L$ derivations are done. In summary, the complexity of the proposed method is $O(pN^2 L^2+NML)$, where $p$ is the number of iterations. 

In comparison, the complexity of the ANM algorithm is $O(N^{3.5} p+N^2L)$, which is obviously slower than our methods in most cases.

\subsection{Number of sources}
One important performance metric of line spectrum estimation method is the number of sources $L$. For the compressed sensing method, the model is determined by the measurement matrix. The number of signal sources, also called sparsity, reflects the degree of freedom of the problem, which directly affects the estimation performance. However, our method does not require a pre-defined measure matrix. Similar to some other existing algorithms [16,22], our method assumes that the number of sources $L$ is known a prior knowledge.  In practical application, we often use the sparseness estimation algorithm [36] to obtain L's estimate as a priori, and then use the line spectrum estimation algorithm to calculate the line spectrum frequency.

When estimating the complex amplitude vector $\mathbf{c}$, our GDLS algorithm process a system of equations with $N$ measurements and $L$ variables to solve. The sparsity $L$ is required to be smaller than $N-1$ to ensure that the system of equations has a solution. The ANM algorithms have similar requirement. In contrast, some CS algorithms, such as ISTA, do not require the prior sparsity information, but require a smaller degree of sparsity with $O(\log(N))$.

\subsection{Resolution}
Frequency spacing which determined the resolution is important in the proposed method . When the frequencies of the two sinusoids are too close, the assumption above to confirm the algorithmic convergence is no longer valid. At this point, the gradient descent method is not necessarily effective. In addition, when the frequency spacing is less than the Rayleigh resolution, the estimated success rate of the OMP algorithm decreases rapidly, and its estimated value is no longer suitable as the initial value of the iteration. Therefore, the theoretical resolution of our GDLS algorithm is the same as the CS method, equal to the Reyleigh resolution which represents as $1/(N-1)$. On the other side, the ANM method ask a fourfold spacing to ensure the SDP problem solving.
\section{Numerical experiments}
In this section, we use simulations to validate our proposed GDLS, compared to the most commonly used OMP algorithm and gridless ANM method. If not specifically stated, the default simulation settings are listed in Table \ref{table1}. In the implementation of the OMP algorithm, the number of grids is set as 1024.
\begin{table}[!t]
	\caption{Default parameters for experiments.}
	\centering
	\label{table1}
		\begin{tabular}{|l|l|}
			\hline
			Size of measurement($N$)	& $16$ \\
			\hline
			Sparsity($L$)	&  $4$\\
			\hline
			Real frequencies ($f_1,…,f_L$)	& $[0.35 0.1 0.67 0.92]$ \\
			\hline
			Real complex amplitude ($c_1,…,c_L$)	&  $[12, 8, 10, 11]$\\
			\hline
			Threshold ($\epsilon$)	& $0.005$ \\
			\hline
		\end{tabular}
	
\end{table}

\subsection{Noise robustness}
\begin{figure}[h]
	\centering
	\includegraphics[width=2.5in]{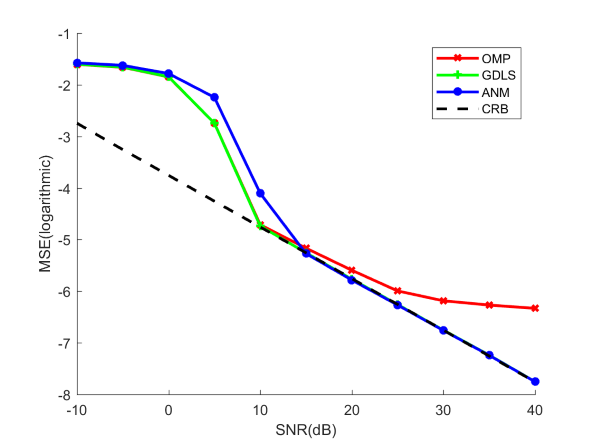}
	\caption{Comparison of MSE performance versus SNR}
	\label{fig3}
\end{figure}

Firstly, we investigate the performance of our method against noise. Fig \ref{fig3} shows the mean square error (MSE) performance of recovered $\mathbf{f}$ versus SNR, compared with OMP and ANM methods. The Cramer-Rao bound (CRB) is also given as a benchmark. When the SNR is high, the MSE performance of our GDLS is close to the ANM algorithm, approaching the CRB. Both of them perform better than the OMP algorithm because its precision is limited by the grid size. When the SNR is low, our GDLS methods performs more similar to OMP and better the ANM algorithm, because the sparse structure of the signal is more stable than the Vander-monde structure. It shows that the proposed method has better noise robustness compared with existing methods.
\subsection{Resolution}
We test the resolving performance by controlling the frequency separation between two sources. In the simulation, we set two sources with the same amplitude and the frequency separation changes from $0.01$ to $0.1$. The success rate is defined as the probability that the estimation error of any frequency is less than a certain threshold $\epsilon$ with $20$dB noise, such as
\begin{equation}\label{key37}
	\left\{\begin{aligned}
		&\max(\vert f_l-\hat{f}_l\vert) < \epsilon,\quad &\mbox{Success}\\
		&else,&\mbox{False}
	\end{aligned}\right.
\end{equation}

\begin{figure}[h]
	\centering
	\includegraphics[width=2.5in]{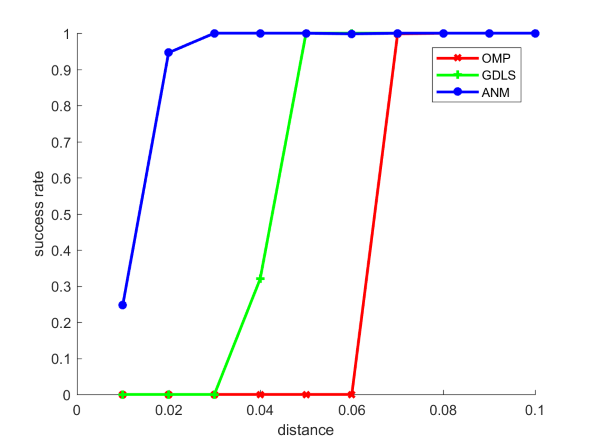}
	\caption{The successful recovery rate versus the minimum frequency separation}
	\label{fig4}
\end{figure}

In the simulation, $\epsilon$ is set to $0.01$. The results are given in Fig \ref{fig4}. We can easily observe that the resolution limit of the GDLS algorithm is about $0.05$, less than the OMP but larger than the ANM algorithm.
\subsection{Impact of the complex amplitude}
Ideally, we want that all the sinusoidal componentss have similar complex amplitudes, otherwise, the sinusoid with lower amplitude will be covered by the noise. On the other side, the phase of complex amplitude will also affect the recovery results. We test the success rate versus the non-uniformity of the complex amplitude and set it of the same mean value and different standard deviations. Two possible phase distributions, zero phase and random phase, are considered.

\begin{figure*}[h]
	\centering
	\subfloat[]{
		\includegraphics[width=2.5in]{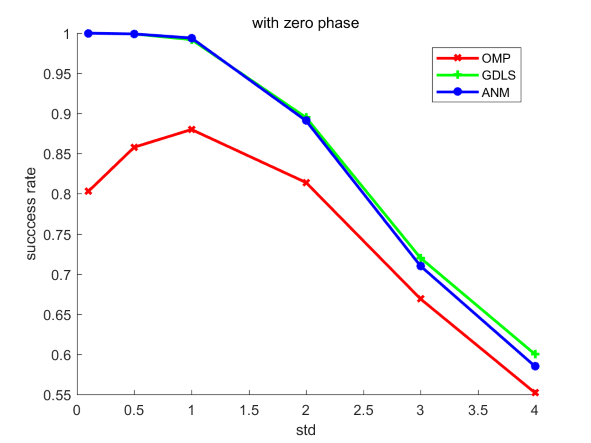}
		\label{subfig5a}
	}
	\hfil
	\subfloat[]{ \includegraphics[width=2.5in]{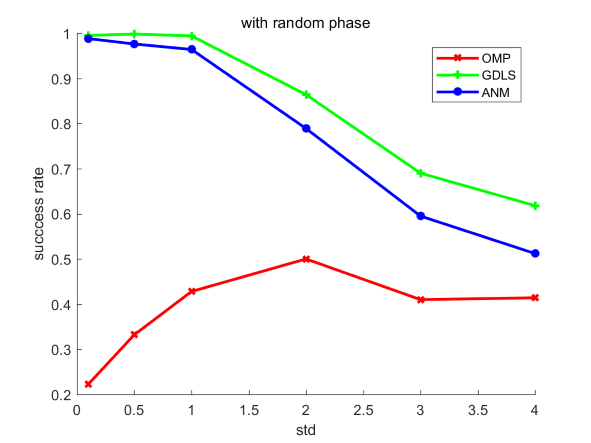}
		\label{subfig5b}}
	\caption{The successful recovery rate versus the Std of amplitude. (a) Real amplitude. (b) Complex amplitude with rando phase.}
	\label{fig5}
\end{figure*}

Fig \ref{fig5} shows that, the success rate of GDLS is generally greater than or at least equal to the other two algorithms.In addition, the GDLS and the ANM algorithm have the best recovery performance for signals with the same amplitude, while the OMP algorithm requires the amplitudes a little different from each other.
\subsection{Running time}
\begin{figure}[h]
	\centering
	\includegraphics[width=2.5in]{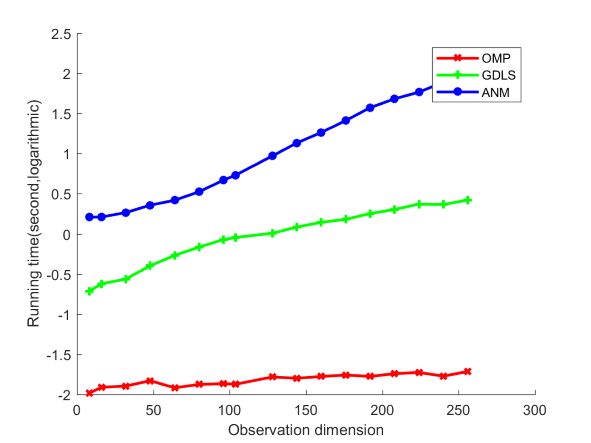}
	\caption{Computing complexity: run time versus N. Time scale: log10 second.}
	\label{fig6}
\end{figure}

Fig \ref{fig6} shows the comparison of the running time of the three algorithms. We tried different $N$, and count the average estimated time of 50 random signals. The experiment is conducted on a personal computer with Intel(R) Core(TM) i5 CPU. It is clear that the proposed method uses much less time than the ANM algorithm. Also, the growth curve of the running time is basically consistent with the analysis results.

\section{Applications in TomoSAR Imaging}

Recall the TomoSAR model provided in section 2.2. Using simulated and real experiment data, the proposed GDLS methods in applied to TomoSAR imaging.

\subsection{Simulation results}

\begin{figure*}[!h]
	\centering
	\subfloat[]{
		\includegraphics[width=2.5in]{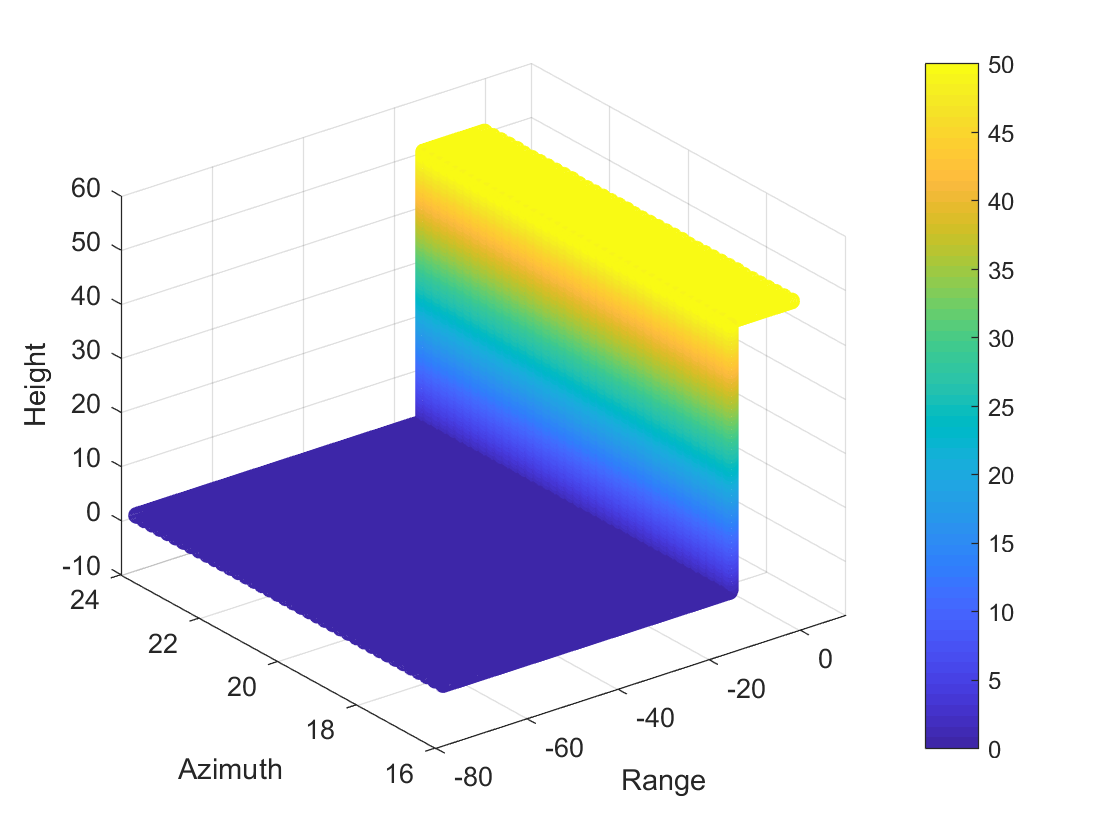}
	}
\hfil
	\subfloat[]{ \includegraphics[width=2.5in]{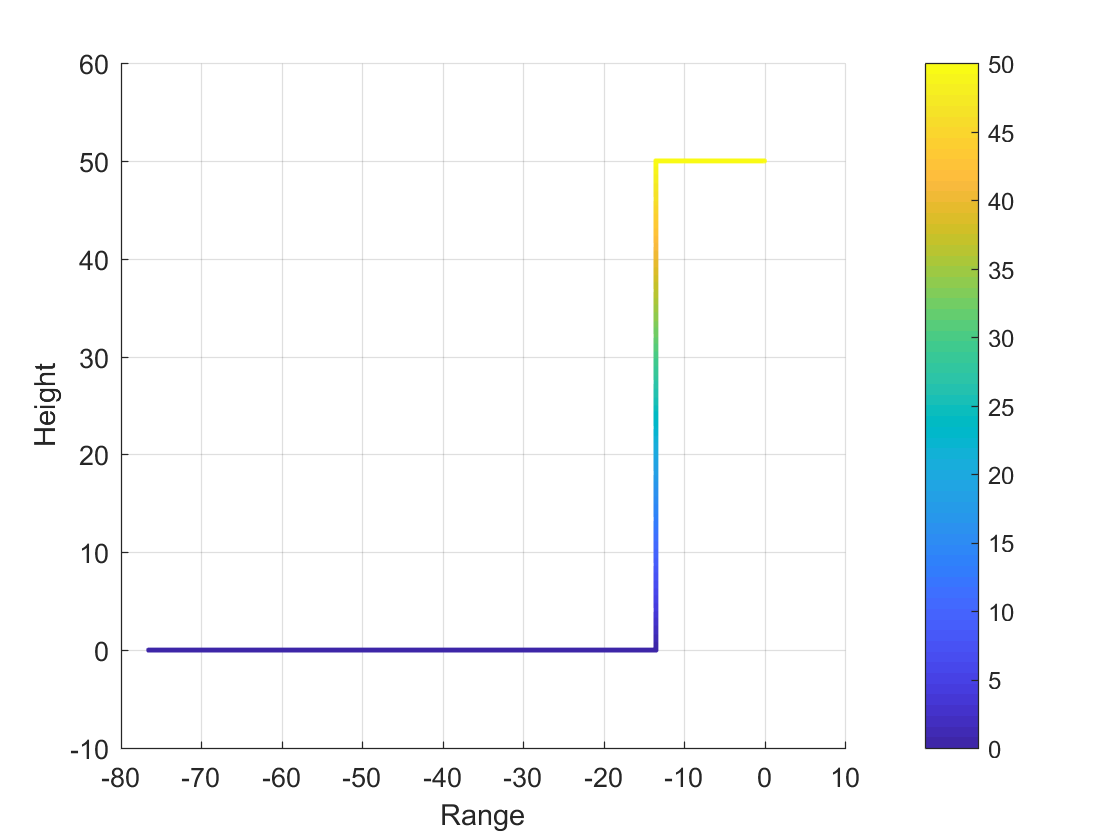}
	}
	\caption{The simulated 3D model of the target building. (a)The 3D model. (b) Slice in the ground height dimension. }
	\label{fig_model}
\end{figure*}

A simulated TomoSAR scenario is set up to test the practical effectiveness of our algorithm. The parameters of the radar system are shown in in Table \ref{table3}. The simulated observation target is a 50 meter high rectangular building, as shown in Fig \ref{fig_model}. There are three possible overlapping targets simulated in the scene--the ground, the floor and the roof. It is assumed that their scattering coefficients have same amplitude and random phase. 
\begin{table}[!t]
	\centering
	\caption{Parameters of the simulated TomoSAR system}
	\label{table3}
		\begin{tabular}{|l|l|}
			\hline
			System frequency	& $14.25$GHz \\
			\hline
			Array baselines	&  $0.084 \times 8$m\\
			\hline
			Range	& $500$m \\
			\hline
			Incident angle	&  $45^{\circ}$\\
			\hline
			SNR & $25$dB\\
			\hline
		\end{tabular}
\end{table}

\begin{figure*}[h]
	\centering
	\subfloat[]{
		\includegraphics[width=2.5in]{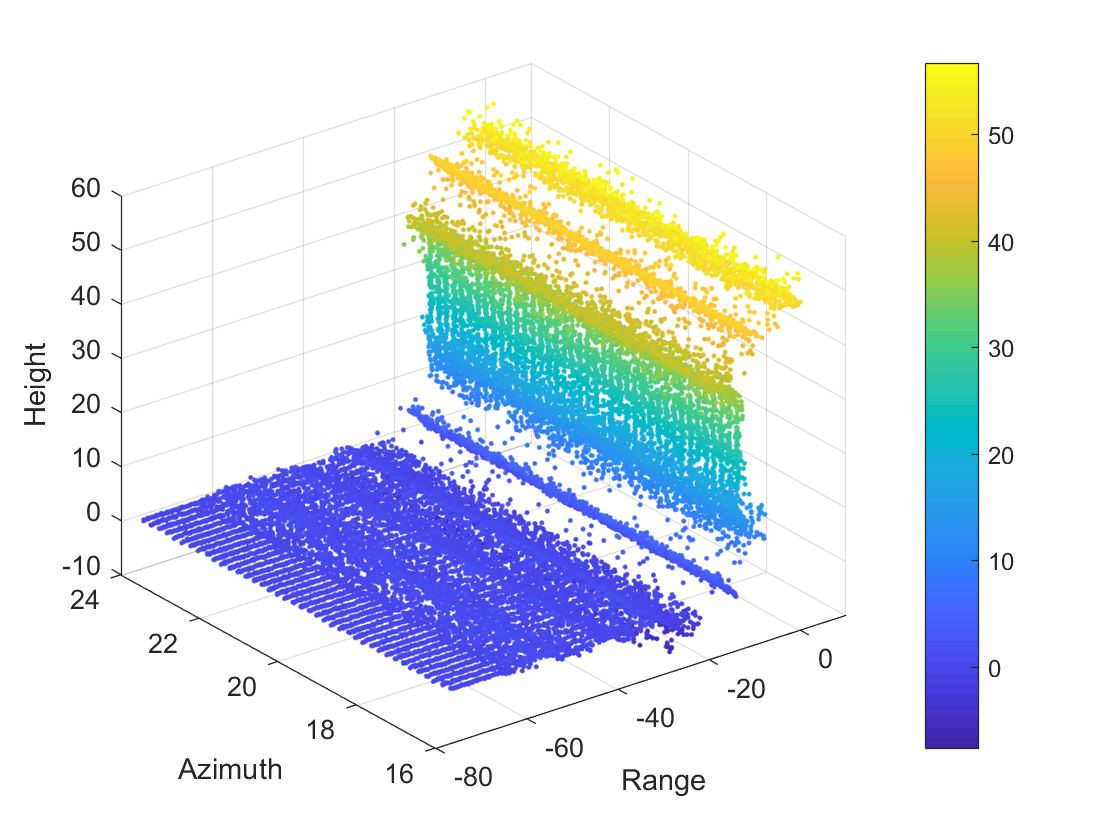}
	}
\hfil
	\subfloat[]{ \includegraphics[width=2.5in]{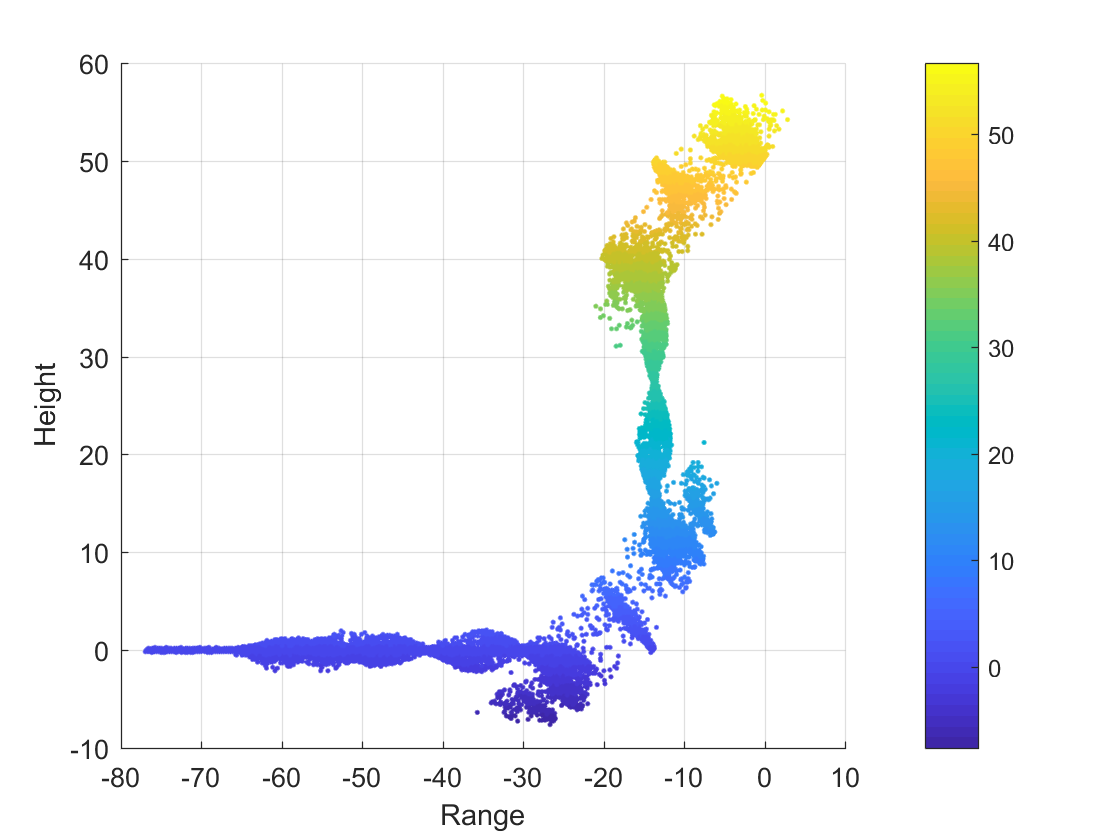}
	}
	\\
	
	\subfloat[]{
		\includegraphics[width=2.5in]{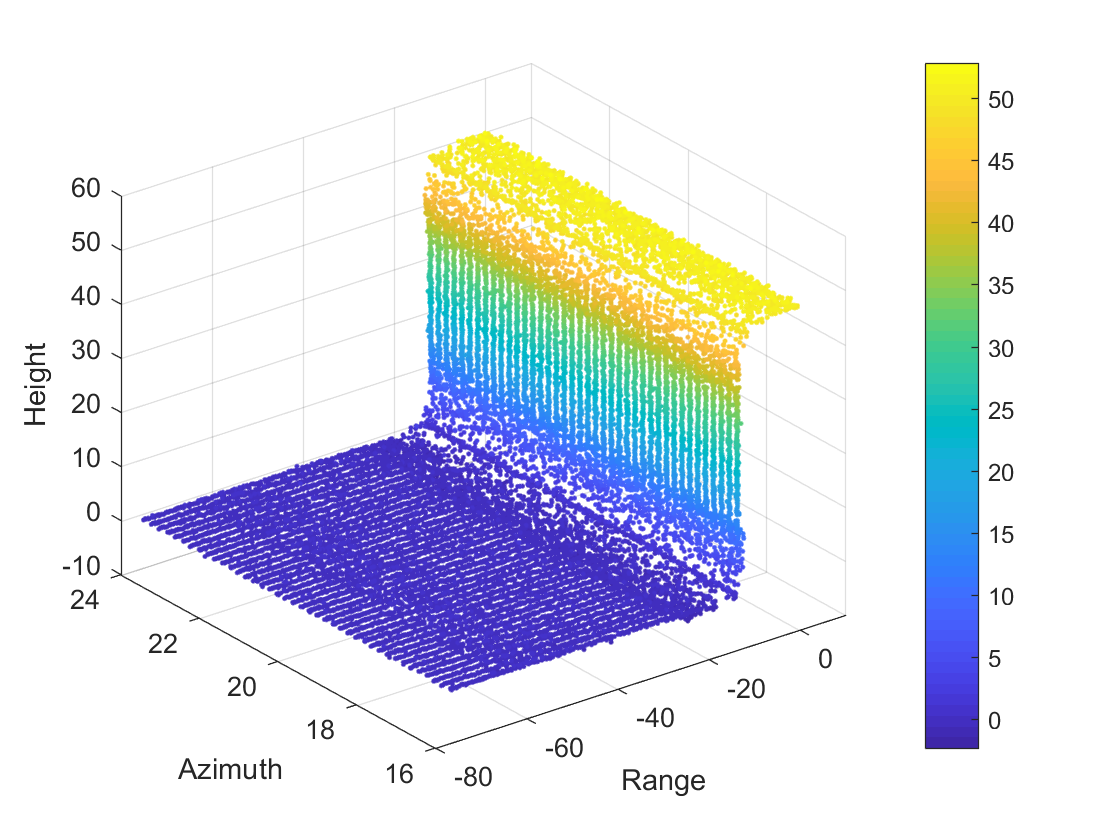}
	}
\hfil
	\subfloat[]{ \includegraphics[width=2.5in]{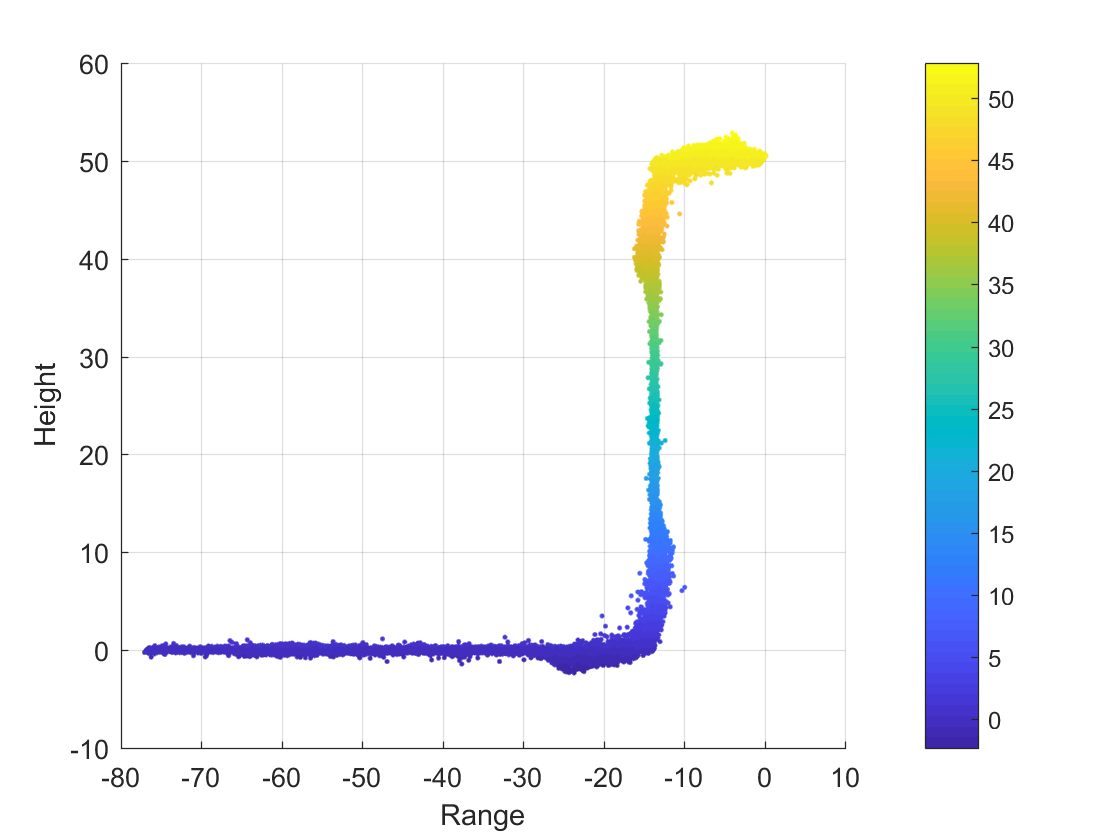}
	}
	\\
	\subfloat[]{
		\includegraphics[width=2.5in]{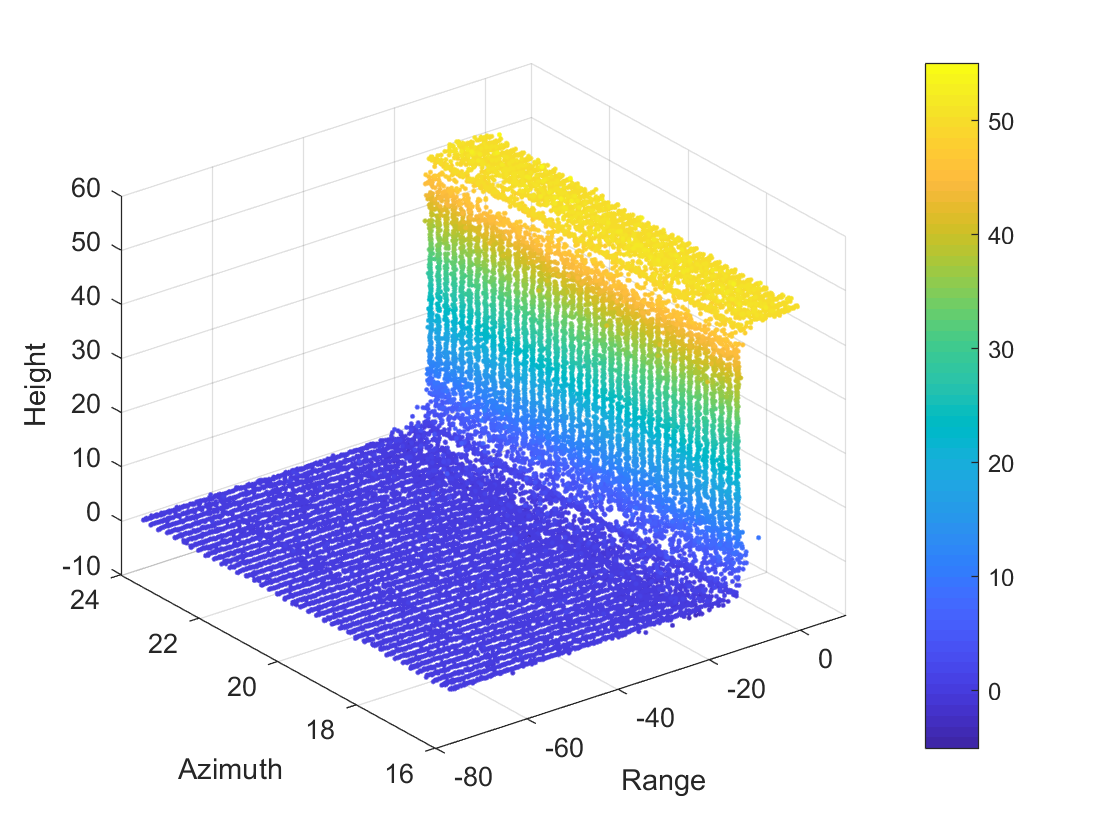}
	}
\hfil
	\subfloat[]{ \includegraphics[width=2.5in]{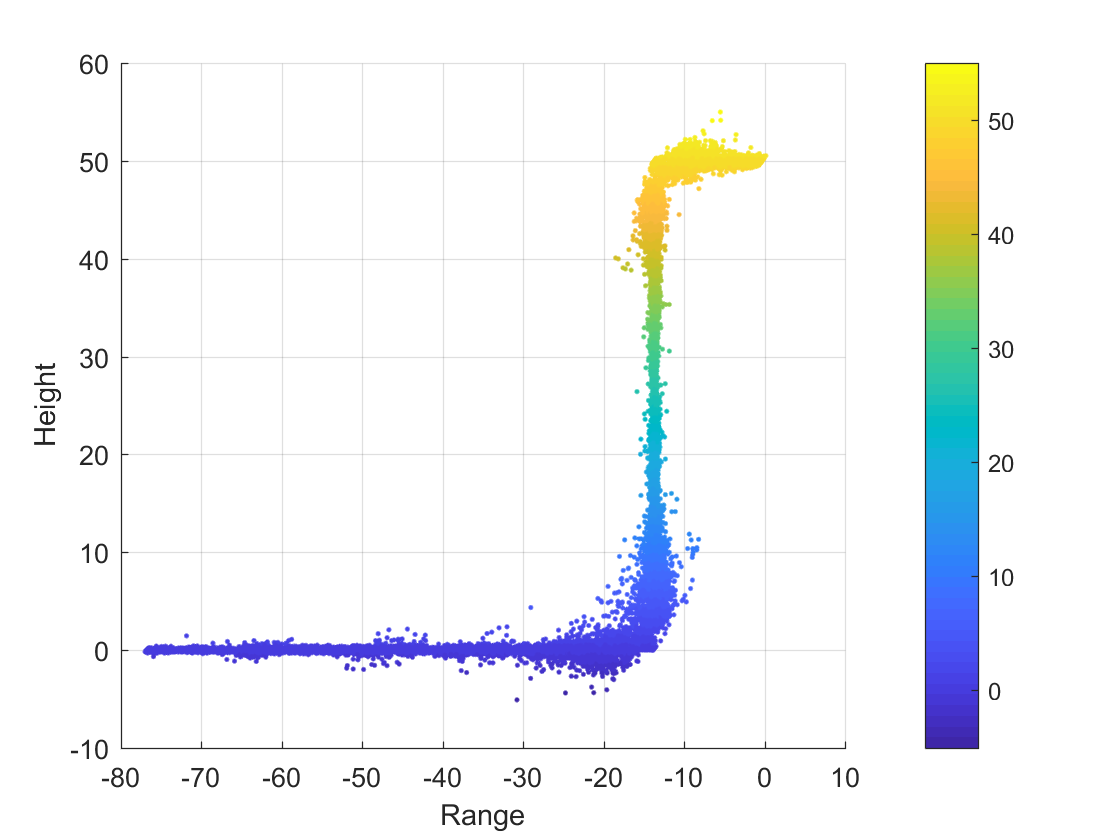}
	}
	
	\caption{The estimated 3d point cloud of the three algorithms. (a),(b) The OMP algorithm. (c),(d) The ANM algorithm. (e),(f) The proposed GDLS algorithm}
	\label{fig_estimate}
\end{figure*}

The three-dimensional point clouds estimated by three algorithms are shown in Fig \ref{fig_estimate}. There are many false target points in the estimation results of the OMP algorithm and the estimation accuracy fluctuates obviously along the change of oblique distance, which means the resolution of elevation direction is poor. By contrast, the gridless GDLS algorithm and the ANM algorithm provide smaller estimation error, and more uniform point cloud. Table \ref{table_est} records the estimation error and computational time of the three algorithms. The GDLS algorithm obtains the estimated point cloud of approximate quality with much shorter time than ANM. The result is consistent with the conclusion of performance analysis above.

\begin{table}[!t]
	\centering
	\caption{Comparision of the three estimated algorithm}
	\label{table_est}
		\begin{tabular}{|c|c|c|c|}
			\hline
			& OMP & ANM & GDLS \\
			\hline
			Time(s) & 5.729157 & 7423.734962 & 849.369583 \\
			\hline
			RMSE(m) & 1.8405 & 0.4839 & 0.5529 \\
			\hline
		\end{tabular}
\end{table}

\subsection{Real data experiments}
\begin{figure}[!h]
	\centering
	\subfloat[]{
		\includegraphics[width=.8\columnwidth]{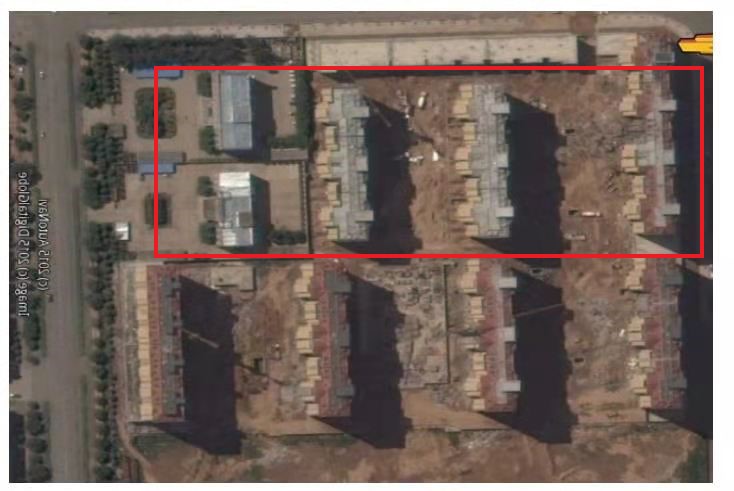}
	}
\hfil
	\subfloat[]{ \includegraphics[width=.8\columnwidth]{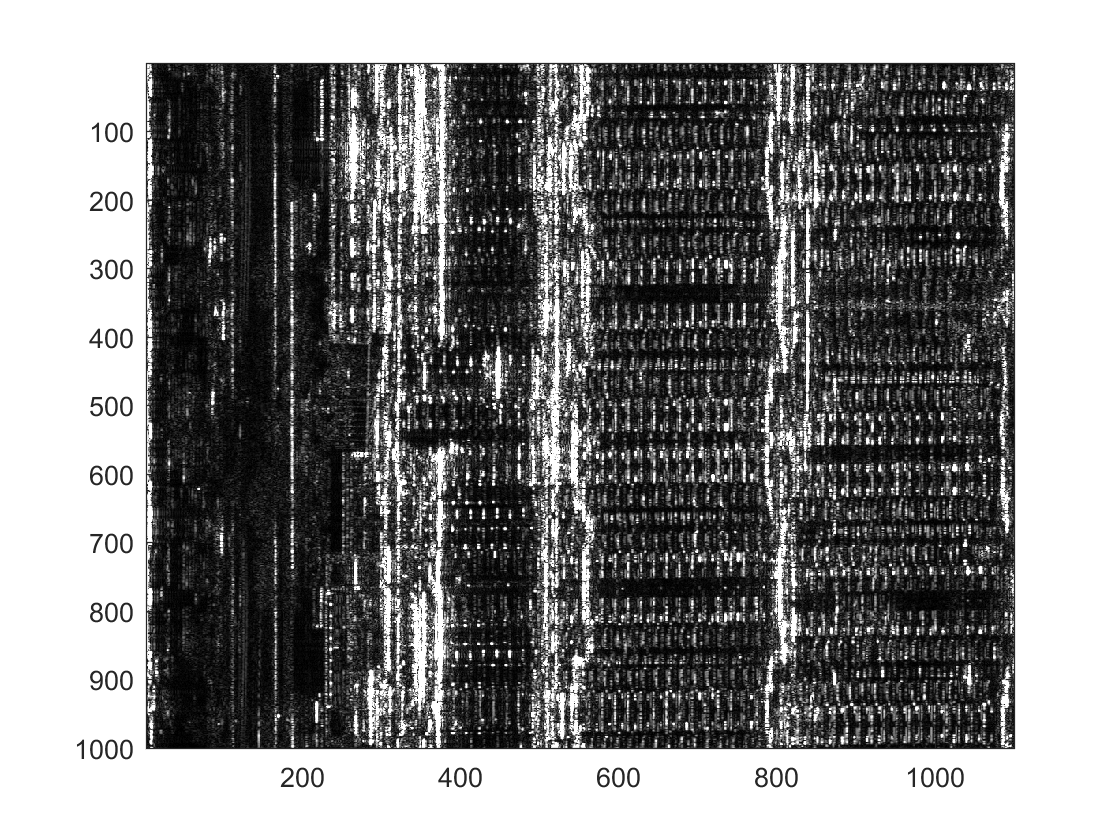}
	}
	\caption{The images of the target scene. (a)The optical image from Google Earth. (b) Imaging result of 2D SAR.}
	\label{fig7}
\end{figure}

We also test the practicability of the proposed algorithm in TomoSAR application with real radar data acquired by Aerospace Information Research Institute, Chinese Academy of Sciences \cite{dataset}. The radar system is an airborne SAR operates at Ku-band. A cross-track antenna array with eight array elements is installed on-board to acquire the TomoSAR data. The experiment settings are given in Table \ref{table2}. The target scene is a housing estate in Yuncheng, Shanxi Province. Fig \ref{fig7} shows the optical image and the SAR imaging result of the target scene. We select five buildings in the target scene, the first two on the left are shorter, and the reminders are basically with the same height. 
\begin{table}[!t]
	\centering
	\caption{System parameters of the airborne array-InSAR}
	\label{table2}

		\begin{tabular}{|l|l|}
			\hline
			System frequency	& $14.25$GHz \\
			\hline
			Array baselines	&  $0.084 \times 8$m\\
			\hline
			Range	& $1184\sim 1304$m \\
			\hline
			Incident angle	&  $24.9\sim 34.5^{\circ}$\\
			\hline
		\end{tabular}
	
\end{table}

\begin{figure}[!h]
	\centering
	\includegraphics[width=2.5in]{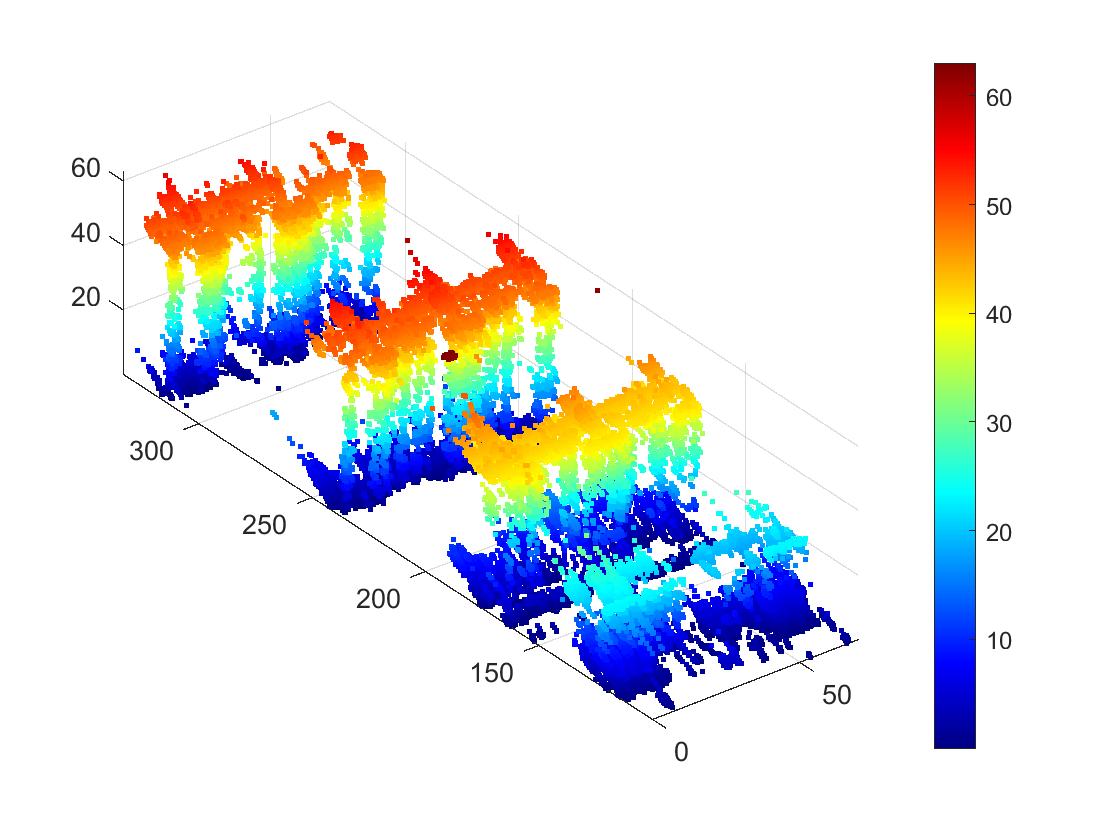}
	\caption{The estimated 3-D imaging point cloud}
	\label{fig8}
\end{figure}

\begin{figure}[!h]
	\centering
	\includegraphics[width=2.5in]{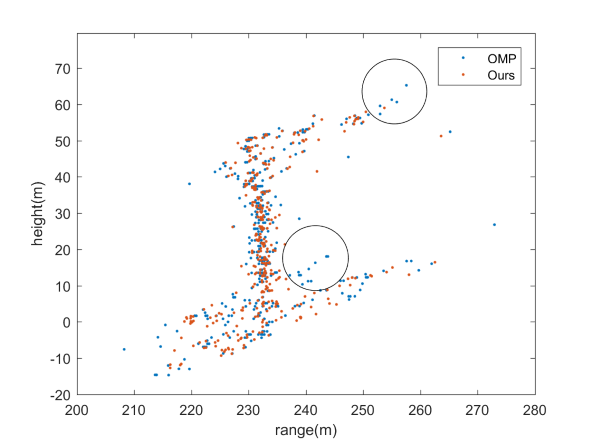}
	\caption{An enlarged comparison between the OMP results and GDLS results.}
	\label{fig9}
\end{figure}

The estimated 3-D imaging point cloud  leveraging the proposed method is given in Fig \ref{fig8}. It can be seen that the GDLS method successfully reconstructed the 3-D information of target scene. The layovered floor and ground can be clearly observed, and the relative heights of the buildings are also correct. An enlarged comparison with the OMP results is shown in Fig \ref{fig9}. The OMP algorithm’s estimation has many incorrect target points in the circled area, while the proposed method does not, which shows the better accuracy of proposed method.

The estimated point clouds on the top and bottom of the building are blurred, mainly due to the insufficient Rayleigh resolution of the system and the interference of multipath scattering. In general, the proposed method is effective in real TomoSAR applications.

\section{Conclusion}
In this paper, we proposed a new line spectrum estimation method by modifying the minimum norm criterion of measurements. The proposed method takes advantage of the mathematical optimization of ML algorithm, while avoids the problem of difficult calculation by optimizing the result of the OMP algorithm. Combining the OMP algorithm and the gradient descent method, the algorithm works for single snapshot case without suffering from the gridding effect. Compared with existing state-of-the-art ANM method, the proposed method has a better performance in terms of noise robustness, resolution and amplitude diversity, while runs faster for more than 1.5 orders. The effectiveness of proposed method in tested in TomoSAR applications.

\vfill

\end{document}